\documentclass[aps,prb,showpacs,preprint,groupedaddress]{revtex4}
\usepackage{graphicx}

\begin{document}


\title{The Magnetic Phase Diagram and the Pressure and Field Dependence of the Fermi Surface in UGe$_2$}


\author{T. Terashima}
\author{T. Matsumoto}
\author{C. Terakura}
\author{S. Uji}
\affiliation{National Institute for Materials Science, Tsukuba, Ibaraki 305-0003, Japan}

\author{N. Kimura}
\author{M. Endo}
\author{T. Komatsubara}
\author{H. Aoki}
\affiliation{Center for Low Temperature Science, Tohoku University, Sendai, Miyagi 980-8578, Japan}

\author{K. Maezawa}
\affiliation{Department of Liberal Arts and Sciences, Toyama Prefectural University, Kosugi, Toyama 939-0398}

\date{\today}

\begin{abstract}
The ac susceptibility and de Haas-van Alphen (dHvA) effect in UGe$_2$ are measured at pressures {\it P} up to 17.7 kbar for the magnetic field {\it B} parallel to the {\it a} axis, which is the easy axis of magnetization.  Two anomalies are observed at {\it B$_x$}({\it P}) and {\it B}$_m$({\it P}) ({\it B$_x$} $>$ {\it B}$_m$ at any {\it P}), and the {\it P}-{\it B} phase diagram is presented.  The Fermi surface and quasiparticle mass are found to vary smoothly with pressure up to 17.7 kbar unless the phase boundary {\it B$_x$}({\it P}) is crossed.  The observed dHvA frequencies may be grouped into three according to their pressure dependences, which are largely positive, nearly constant or negative.  It is suggested that the quasiparticle mass moderately increases as the boundary {\it B$_x$}({\it P}) is approached.  DHvA effect measurements are also performed across the boundary at 16.8 kbar.
\end{abstract}


\maketitle

The recent discovery of superconductivity in the itinerant-electron ferromagnet UGe$_2$ by Saxena {\it et al}. has aroused much excitement.\cite{Saxena00}  This could be the superconductivity of the type that has long been sought for, i.e., the superconductivity mediated by ferromagnetic spin fluctuations.\cite{Coleman00NV}  However, the superconductivity in UGe$_2$ does not rigorously conform to previous theoretical expectations in that it occurs only in the ferromagnetic phase.  It is theoretically anticipated that, as a ferromagnetic transition is continuously suppressed down to absolute zero, spin fluctuations are enhanced and may lead to magnetically mediated superconductivity on both ferromagnetic and paramagnetic sides of the quantum critical point.\cite{Fay80,Roussev01}  On the one hand, the peculiarity of the superconductivity in UGe$_2$ may be attributed to some particular features of the compound, as further discussed below.  On the other hand, the fact that the superconductivity in the itinerant-electron ferromagnet ZrZn$_2$ also disappears when the ferromagnetism vanishes (Ref.~\onlinecite{Pfleiderer01}) may suggest that ferromagnetic order is a prerequisite for the superconductivity in these compounds.  Answering this essential question will require detailed understanding of the electronic structure, to which the present work is intended to contribute.

The Curie temperature {\it T}$_C$ in UGe$_2$, being 52 K at ambient pressure,\cite{Menovsky83} decreases with pressure and vanishes at the critical pressure {\it P}$_c$ $\sim$ 16 kbar.\cite{Saxena00, Nishimura94, Oomi98, Tateiwa01JPCM, Huxley01}  It has been suggested that the ferromagnetic transition at pressures near {\it P}$_c$ is first order.\cite{Huxley00, Terashima01}  An additional anomaly is found at {\it T$_x$} ($<$ {\it T}$_C$) in the ferromagnetic phase;\cite{Oomi98, Tateiwa01JPCM, Huxley01, Tateiwa01JPSJ} the temperature derivative of resistivity shows a broad peak at {\it T$_x$}, and magnetization increases below {\it T$_x$}.  The characteristic temperature {\it T$_x$} also decreases with pressure and appears to reach absolute zero at {\it P$_x$} $\sim$ 12-13 kbar.  The origin of the {\it T$_x$} anomaly is not yet clear.  It has been proposed that the anomaly is due to the formation of coupled charge- and spin-density-waves.\cite{Saxena00, Huxley01, Watanabe01}  The superconductivity appears below 1~K in a pressure range $\sim$10-16 kbar.\cite{Saxena00, Tateiwa01JPCM, Huxley01, Terashima01, Bauer01}  The transition temperature is highest at pressures near {\it P$_x$}.  This leads to the conjecture that the superconductivity is mediated by fluctuations associated with the second-order quantum critical point at {\it P$_x$} rather than {\it P}$_c$.\cite{Saxena00, Huxley01, Watanabe01}  It is therefore of importance to clarify the origin of the {\it T$_x$} anomaly and its influence on quasiparticle properties.

The magnetic response of UGe$_2$ is extremely anisotropic; at 4.2 K, the {\it b}-axis magnetization is less than 15$\%$ of the {\it a}-axis one even in a field of 35 T.\cite{Menovsky83}  In our previous de Haas-van Alphen (dHvA) effect measurements,\cite{Terashima01} the magnetic field was applied parallel to the {\it b} axis, a hard axis of magnetization, and hence the Fermi surface and related properties determined in those measurements are virtually those at zero magnetic field.  In this work, we apply the field along the easy {\it a} axis.  The field along the {\it a} axis induces two phase transitions at high pressures, which are intimately related to the {\it T$_x$} anomaly and the ferromagnetic transition.  We determine the pressure vs field phase diagram by measuring ac susceptibility and study the Fermi surface and quasiparticle mass as functions of pressure and of field via the dHvA effect.

The single crystal used in the present measurements was grown by the Czochralski method and subsequently annealed at 1100$^{\circ}$C for 110 hours under ultra-high vacuum.  The residual resistivity ratio along the {\it a} axis is about 200.  Hydrostatic pressures {\it P} up to 17.7 kbar were produced by a BeCu/NiCrAl clamped piston-cylinder cell with Daphne 7373 oil (Idemitsu Co. Ltd., Tokyo) as a pressure-transmitting medium, and ac susceptibility, the oscillatory part of which comes from the dHvA effect, was measured with a pick-up coil (see Ref.~\onlinecite{Terashima01} for details).  Since the sample is ferromagnetic, the magnetic field {\it B} inside the sample differs from the applied field {\it B}$_{appl}$; {\it B} = {\it B}$_{appl}$ + $\mu$$_0$(1-{\it N}){\it M}, where {\it N} and {\it M} are the demagnetization factor and magnetization, respectively.  We estimated {\it N} to be 0.1 from the sample shape and {\it M} from data in Ref.~\onlinecite{Tateiwa01JPSJ}.

\begin{figure}
\includegraphics[width=8cm]{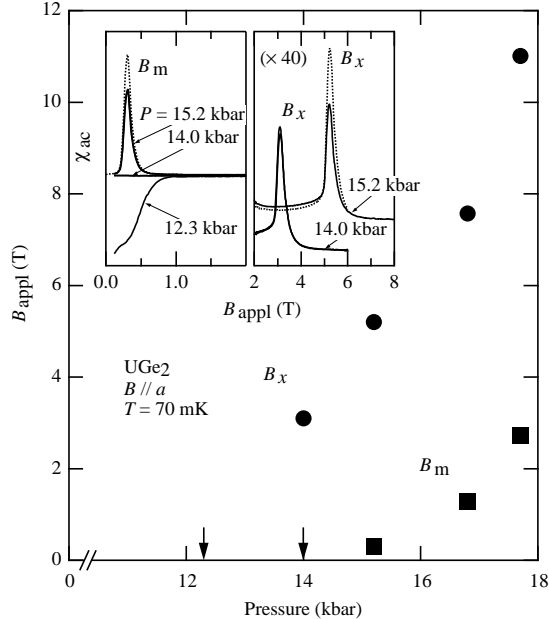}
\caption{\label{fig:1}The inset shows the ac susceptibility along the {\it a} axis for selected pressures.  The anomaly fields {\it B}$_m$ and {\it B$_x$} are indicated.  The measurement temperature is 0.07 K, except the dotted curves measured at 1.1 K.  Up- and down-field-sweep data are superimposed near {\it B}$_m$ for 15.2 kbar and near {\it B$_x$} for 14.0 kbar to show the absence of hysteresis at those anomalies.  The vertical scale of the right panel is expanded by the factor of 40.  Since the balance of a pick-up coil slightly varies from pressure to pressure, and since this effect is not corrected, a vertical shift between the 14.0 and 15.2 kbar data has no significance.  The main panel shows {\it B}$_m$ and {\it B$_x$} as functions of pressure.  The two arrows indicate that {\it B}$_m$ and {\it B$_x$} are absent at the respective pressures.}
\end{figure}

\begin{figure}
\includegraphics[width=8cm]{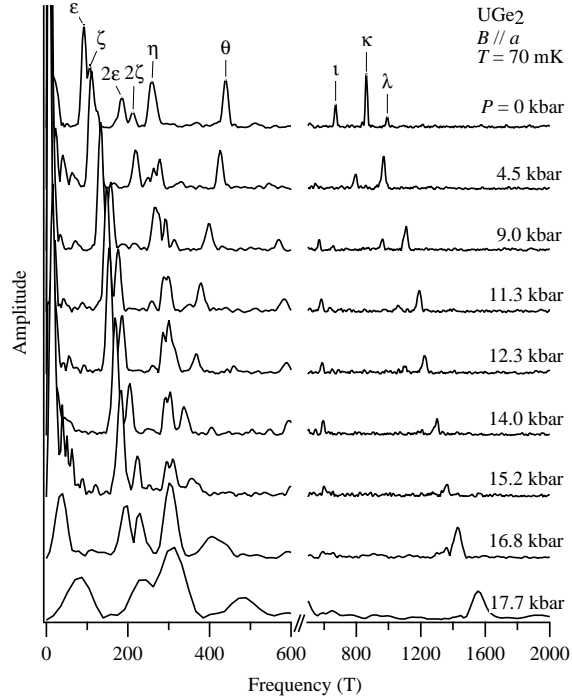}
\caption{\label{fig:2}Fourier spectra of dHvA oscillations along the {\it a} axis in UGe$_2$ as a function of pressure.  DHvA frequencies, or orbits, are labeled by Greek letters.  Each spectrum is arbitrary scaled for clarity.  The data window for the Fourier transformations is approximately from {\it B}$_{appl}$ = 5 to 18 T for pressures up to 15.2 kbar, while the window is narrowed for higher pressures to avoid the anomaly at {\it B$_x$}; {\it B}$_{appl}$ = 8.2-17.6 T for 16.8 kbar and 11.6-17.6 T for 17.7 kbar.  Because of the narrower windows, the frequency resolution is deteriorated for these pressures. }
\end{figure}

\begin{figure}
\includegraphics[width=8cm]{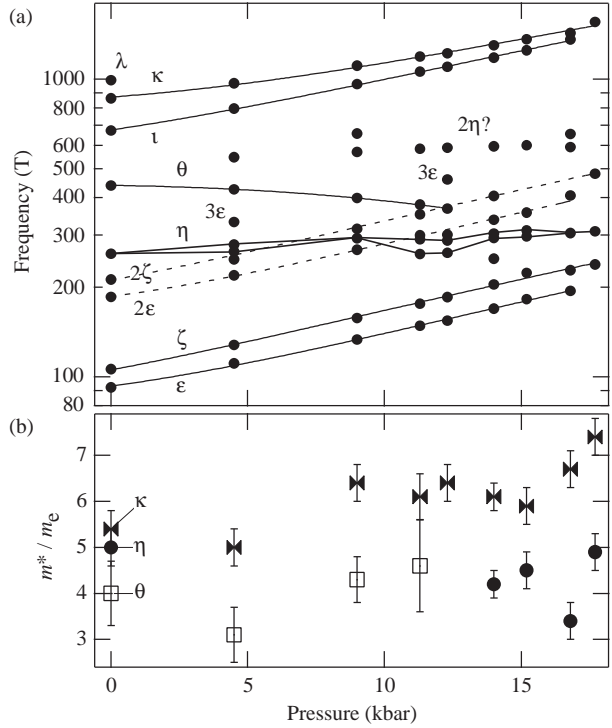}
\caption{\label{fig:3}Pressure dependence of (a) the dHvA frequencies and (b) the effective masses associated with the orbits $\eta$, $\theta$ and $\kappa$ (in the units of free electron mass {\it m}$_e$).  The masses were determined from the temperature dependence of oscillation amplitudes as usual.  The field span of oscillation data used in the mass determination is approximately from {\it B}$_{appl}$ = 11 (11.5 for 17.7 kbar) to 18 T.  Since the windows are narrower than those used for the spectra in Fig.~\ref{fig:2}, all the frequencies in Fig.~\ref{fig:2} are not resolved.  The figure shows the masses only for the frequencies that are well resolved at pressures of a wide range.}
\end{figure}

\begin{figure}
\includegraphics[width=8cm]{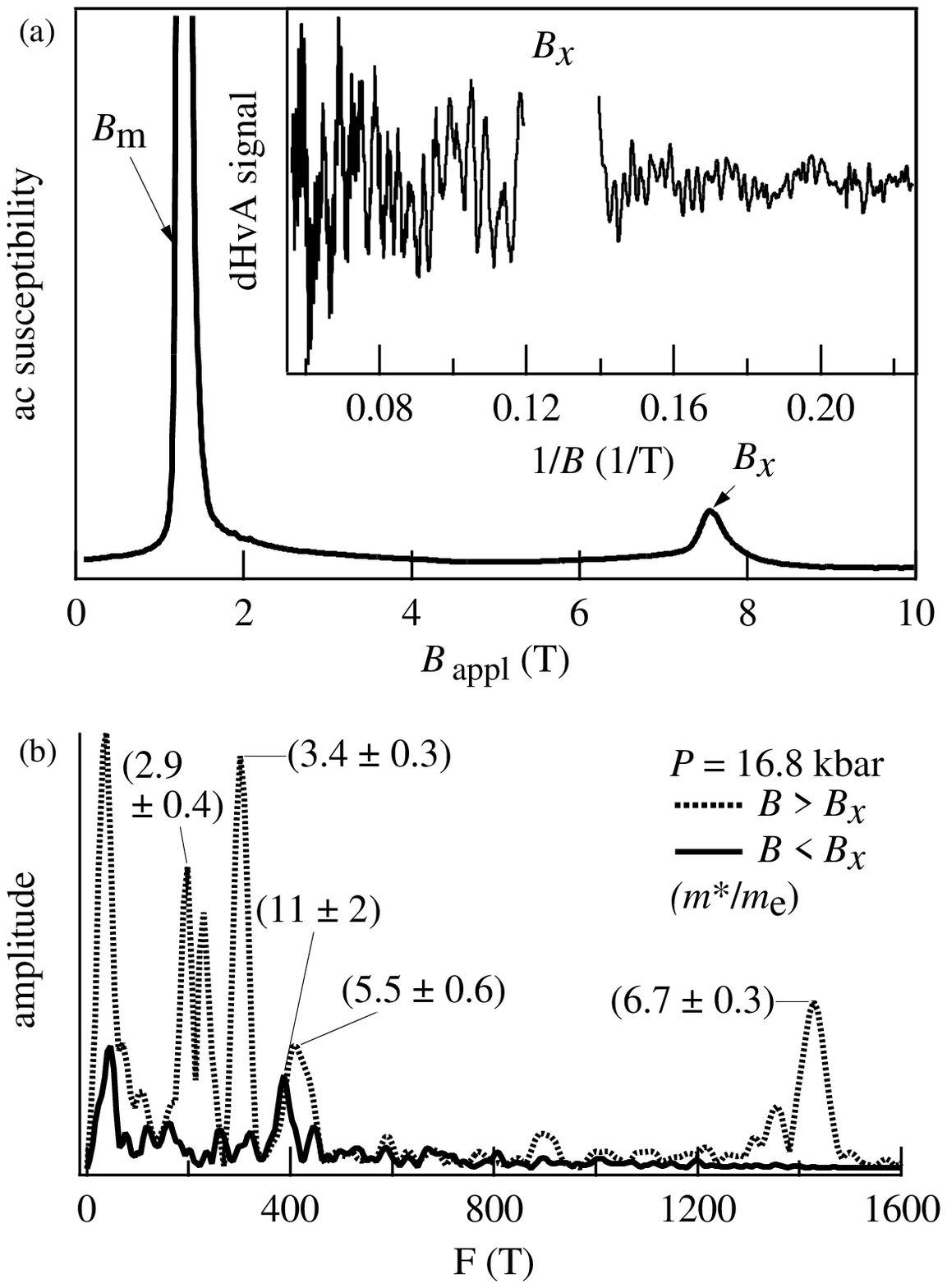}
\caption{\label{fig:4}(a) AC susceptibility at 16.8 kbar.  The inset shows dHvA oscillations below and above the anomaly field {\it B$_x$}.  Measurement conditions were different, so that the amplitudes of oscillations can not directly be compared between below and above {\it B$_x$}.  (b) Fourier spectra of oscillations below and above {\it B$_x$}.  The quasiparticle effective masses are shown for some orbits in the parentheses.}
\end{figure}

The inset of Fig.~\ref{fig:1} shows ac susceptibility for selected pressures.  The ac susceptibility at 12.3 kbar exhibits a superconducting diamagnetic signal at low fields, while those at 14.0 and 15.2 kbar show one and two anomalies, respectively.  The anomaly fields {\it B}$_m$ and {\it B$_x$} are shown as functions of pressure in the main panel.

The absence of diamagnetic signals at pressures other than 12.3 kbar could indicate that the pressure range for the superconductivity in this particular sample is extremely narrow.  However, we suspect that diamagnetic signals at other pressures are simply suppressed below the detection limit by the ac excitation field of 0.62 mT applied along the {\it a} axis.  Actually, Saxena {\it et al}. used one order-of-magnitude smaller excitation fields to observe appreciable diamagnetic signals at $\sim$15 kbar.\cite{Saxena00}

The anomaly at {\it B}$_m$ corresponds to what Huxley {\it et al} attributed to a metamagnetic transition.\cite{Huxley00}  In the framework of itinerant-electron metamagnetism,\cite{Moriya86, Yamada93} the transition is expected to be a first-order one from the paramagnetic state to a polarized state where up- and down-spin electron energy bands are split as they are in the ferromagnetic state.

Since the susceptibility peak at {\it B}$_m$ is fairly large, the possibility that it is due to a first-order transition is not excluded.  The absence of hysteresis may be an indication that it is too small to be observed.  Although the peak width, $\sim$0.1 T at the half maximum, may appear broad, it may be explained by tiny pressure variation ($\sim$0.3$\%$) over the sample.  The suppression of the peak height at lower temperatures might indicate that domain-wall motion is involved in the transition process, as is the case with a first-order transition.

The pressure {\it P}$_{c0}$ where {\it B}$_m$ reaches zero is in between 14.0 and 15.2 kbar (Fig.~\ref{fig:1}), which is consistent with  {\it P}$_{c0}$ of $\sim$14.4 kbar reported by Kobayashi {\it et al}.\cite{Kobayashi01}  On the other hand, the critical pressure {\it P}$_c$ where the ferromagnetism vanishes has been reported to be $\sim$16 kbar,\cite{Saxena00, Huxley01} and in fact we have located it between 15.4 and 17.6 kbar in previous measurements on a different sample.\cite{Terashima01}  The discrepancy between {\it P}$_{c0}$ and {\it P}$_c$ might be due to sample dependence and/or error in pressure determination, which is estimated to be $\sim$$\pm$0.3 kbar in our case.  However, we note that it may indicate the existence of a narrow pressure region, {\it P}$_{c0}$ $<$ {\it P} $<$ {\it P}$_c$, where ferromagnetic order exists at zero field, and a metamagnetic transition is observed in fields.  Similar observations that {\it P}$_{c0}$ $<$ {\it P}$_c$ were also reported for some itinerant-electron metamagnets, e.g., Y(Co$_{1-{\it x}}$Al$_{\it x}$)$_2$ (Ref.~\onlinecite{Goto94}) and UCoAl$_{1-{\it x}}$Ga$_{\it x}$.\cite{Andreev98}  It is, however, questionable whether the ferromagnetism and metamagnetism can microscopically coexist.  In this relation, it is interesting to note a recent report by Motoyama {\it et al}.,\cite{Motoyama01} in which the authors have argued that, when the pressure is increased in the pressure range of the superconductivity, the ferromagnetism in UGe$_2$ may become spatially inhomogeneous.

The susceptibility peak at {\it B$_x$} is so small that it is not a first-order phase transition (note that the vertical scale for the right panel of the inset to Fig.~\ref{fig:1} is expanded by the factor of 40).  The peak height decreases with temperature, the origin of which temperature dependence is not clear.  The anomaly field {\it B$_x$} increases with pressure and appears to be zero at {\it P$_x$} ($\sim$ 12-13 kbar) (Fig.~\ref{fig:1}).    Huxley {\it et al}. previously found the same pressure dependence of {\it B$_x$} and argued that the magnetic field along the {\it a} axis shifted the line {\it T$_x$}({\it P}) in a {\it P}-{\it T} plane to higher pressures.\cite{Huxley01}  Tateiwa {\it et al}. gave a clear support to this suggestion by measuring magnetization vs temperature curves in fields at a constant pressure slightly higher than {\it P$_x$};  the curve measured at the lowest field does not show any sign of the {\it T$_x$} anomaly down to the lowest temperature investigated, while curves measured at higher fields exhibit rapid increase in magnetization, an indication of the {\it T$_x$} anomaly, at temperatures that increase with field.\cite{Tateiwa01JPSJ}  The interpretation of {\it B$_x$} may be rephrased in a way that is more relevant to Fig.~\ref{fig:1}; i.e., the {\it T$_x$} anomaly occurs at finite temperatures on the left side of the line {\it B$_x$}({\it P}), while it does not down to zero temperature on the right side.

Before presenting dHvA data, we here mention two main results of the previous {\it b}-axis dHvA measurements.\cite{Terashima01}  Firstly, we have found that the Fermi surface discontinuously changes as {\it P}$_c$ is crossed.  Secondly, the quasiparticle mass is enhanced near {\it P$_x$}; the mass associated with a large orbit, $\beta$, being 12~{\it m}$_e$ at ambient pressure, gradually increases to 16~{\it m}$_e$ at 11.9 kbar, then suddenly jumps to 39~{\it m}$_e$ at 12.9 kbar, {\it m}$_e$ being the free electron mass.

Figure~\ref{fig:2} shows the Fourier spectra of dHvA oscillations for the field along the {\it a} axis as a function of pressure.  Note that, for pressures where the {\it B$_x$} anomaly is observed, only oscillation data above {\it B$_x$} were Fourier-transformed.  Figure~\ref{fig:3} summarizes dHvA frequencies and effective masses as functions of pressure.  The frequencies and masses at ambient pressure agree well with a previous report.\cite{Satoh92}

Figures~\ref{fig:2} and \ref{fig:3} clearly indicate that the Fermi surface and quasiparticle mass smoothly vary without any discontinuity from 0 to 17.7 kbar.  This is in sharp contrast to the {\it b}-axis results.  The difference is easily explicable in terms of the phase diagram in Fig.~\ref{fig:1}.  As mentioned in the introduction, the {\it b}-axis measurements are virtual zero-field measurements, and hence {\it P$_x$} and {\it P}$_c$ were indeed crossed in the course of the measurements.  On the other hand, the results shown in Figs.~\ref{fig:2} and \ref{fig:3} were obtained along a path that does not intersect the boundary {\it B$_x$}({\it P}).

The dHvA frequencies may be categorized into three according to their pressure dependence [Fig.~\ref{fig:3}(a)].  (1) $\epsilon$, $\zeta$, $\iota$, and $\kappa$ rapidly increase with the pressure coefficient dln{\it F}/d{\it p} of 25-40 x 10$^{-3}$ kbar$^{-1}$, (2) $\theta$ decreases with pressure, and (3) $\eta$ stays nearly constant.  These differences in behavior would be valuable in assigning the frequencies to orbits if band-structure calculations under high pressures became available.

Although the pressure dependence of the effective masses is not very appreciable, a gradual increase, $\sim$40$\%$ from 0 to 17.7 kbar, may be seen for $\kappa$ [Fig.~\ref{fig:3}(b)].  We also found a faint tendency that the masses associated with $\eta$ at 16.8 kbar and $\kappa$ at 17.7 kbar increase as the field is decreased down to within $\sim$2 T of {\it B$_x$}, though the magnitudes of those variations are nearly comparable to the error in the mass determination ($\sim$$\pm$20$\%$) and are left to be determined in more precise measurements.  These observations indicate that the quasiparticle mass moderately increases as the boundary {\it B$_x$}({\it P}) is approached from the left side in Fig.~\ref{fig:1}.  This is consistent with the modest increase in the mass (before the jump) observed in the {\it b}-axis measurements.

It is then interesting to see how the mass changes across the boundary {\it B$_x$}({\it P}).  Figure~\ref{fig:4}(a) shows the ac susceptibility at 16.8 kbar.  As the inset shows, dHvA oscillations are visible both below and above {\it B$_x$}.  Figure~\ref{fig:4}(b) shows the Fourier transforms of the oscillation data below and above {\it B$_x$}, and masses for orbits.  Several frequencies are resolved for {\it B} $>$ {\it B$_x$} (the dotted curve), but the associated masses are 6.7~{\it m}$_e$ at most.  On the other hand, only one frequency is visible for {\it B} $<$ {\it B$_x$} (the solid curve), and the associated mass is 11~{\it m}$_e$.  That is, despite the fact that frequencies with heavy mass is easier to observe at higher fields, the mass of any frequency that is seen above {\it B$_x$} is lighter than the mass of the single frequency that is detected below {\it B$_x$}.  This can easily be understood if we assume, based on the mass jump near {\it P$_x$} found in the {\it b}-axis measurements, that the quasiparticle mass is considerably enhanced as the boundary {\it B$_x$}({\it P}) is crossed to the right (in this case, to the low-field side).  Results of resistivity measurements by Kobayashi {\it et al}. are in favor of this assumption; the quadratic temperature coefficient of resistivity determined as a function of magnetic field at 16.7 kbar ($>$ {\it P$_x$}) is larger below {\it B$_x$} than above.\cite{Kobayashi01}
 
In summary, we have determined the {\it P-B} phase diagram of UGe$_2$, which comprises the two phase boundaries {\it B$_x$}({\it P}) and {\it B}$_m$({\it P}).  While the anomaly at {\it B$_x$} is not of first order, that at {\it B}$_m$ may be of first order.  We have pointed out the possibility that the pressure {\it P}$_c0$ where {\it B}$_m$ reaches zero is slightly lower than {\it P}$_c$.  Together with the recent suggestion that the ferromagnetism may be inhomogeneous in the pressure range of the superconductivity,\cite{Motoyama01} this seems to deserve further investigations.  We have shown that the Fermi surface and quasiparticle mass continuously vary with pressure up to 17.7 kbar on the low-pressure/high-field side of the boundary {\it B$_x$}({\it P}).  This is in sharp contrast with the previous {\it b}-axis results.  The dHvA frequencies may be grouped into three according to the rate of the pressure variation, which would be helpful in assigning each frequency to an orbit on the Fermi surface.  The mass associated with the frequency $\kappa$ shows moderate increase of $\sim$40$\%$ from 0 to 17.7 kbar.  We have also examined the variation of the mass across the boundary {\it B$_x$} at 16.8 kbar.  The result seems consistent with the mass enhancement increasing below {\it B$_x$}.  Our results as a whole suggest that changes in quasiparticle properties across the critical pressures {\it P$_x$} and {\it P}$_c$ may conveniently be revealed by studying those properties as functions of field (in the direction of the {\it a} axis) at high pressures.

\begin{acknowledgments}
Work at Tohoku University was supported by a Grant-in-Aid for Scientific Research of MEXT Japan.
\end{acknowledgments}

\end{document}